\documentclass[showpacs]{revtex4-1}
\usepackage{epsfig}
\usepackage[dvips]{color}
\def\beq{\begin{equation}}
\def\eeq{\end{equation}}
\def\beqa{\begin{eqnarray}}
\def\eeqa{\end{eqnarray}}

\def\GeV{\nobreak\,\mbox{GeV}}

\def\pli{p^\prime}

\usepackage{color}
\usepackage{ulem}

\begin{document}
\title{$B_s^* B K $ vertex from QCD sum rules}

\author{A. Cerqueira Jr.,  B. Os\'orio Rodrigues}
\affiliation{Instituto de F\'{\i}sica, Universidade do Estado do Rio de 
Janeiro, Rua S\~ao Francisco Xavier 524, 20550-900, Rio de Janeiro, RJ, Brazil. }

\author{M. E. Bracco}
\affiliation{Faculdade de Tecnologia, Universidade do Estado do Rio de Janeiro, Rod. Presidente Dutra Km 298, P\'olo Industrial, 27537-000 , Resende, RJ, Brazil.}

\begin{abstract}

The form factors  and the coupling constant of the 
$B_s^* B K$ vertex are calculated using the QCD sum rules method. 
Three point correlation functions are computed considering both $K$ and $B$ 
mesons off-shell and, after an extrapolation of the QCDSR results, we obtain the coupling constant of the vertex.
We study the uncertainties in our result by calculating
a third form factor obtained when the $B^*_s$ is the off-shell meson, considering other acceptable structures and computing the variations of the sum rules' parameters. The form factors obtained have different 
behaviors but their simultaneous extrapolations reach to the same value of the coupling constant $g_{B_s^* B K}=10.6 \pm 1.7$. We compare our result with other theoretical estimates.    
\end{abstract}


\maketitle

\section{Introduction}

  Hadronic three-meson vertex functions or form factors are basic inputs to 
phenomenological theories of  nuclear processes. They play an important role in cross section calculation, which may vary in a few orders of magnitude if they are considered. 
The form factor depends on the momentum $Q^2$ and some parameters.  As usual, the form factor is parametrized by {\it{ad hoc}} function, which is not unique and can be Gaussian, exponential or monopolar. From the parametrization, we can extract a cutoff parameter $\Lambda$ which is associated with the sharpness of the form factor. Therefore, knowing its precise functional form and its cutoff parameter are essential. 

Some years ago, we started a program to compute form factors with the QCD sum rules (QCDSR)
\cite{SVZ}. Our group has developed a method which leads to less ambiguities in the determination of the coupling constant. The method consists of computing two form factors of the same three meson process: one form factor is evaluated when the heavy meson is off-shell and the other form factor is obtained when the light meson is off-shell. By extrapolating both form factors to the momentum $Q^2 $ equal to $ - m_T^2$, where $m_T^2$ is the mass of the off-shell particle, we obtain the coupling constant of the vertex. After our pioneer $D D \rho$ work \cite{bclnn01}, we concluded that when the form factor tends to be harder as a function  of $Q^2$,the off-shell meson in the vertex is heavy, which means the cut-off parameter is larger.
A persistent study of vertices involving charmed mesons ($D^* D \pi$ \cite{nnbcs00,nnb02},  $D D J/\psi$ \cite{nns02}, $D^* D J/\psi$ \cite{smnn04}, $D^* D^* \pi$ \cite{wang,cdnn05}, $D^* D^* J/\psi$ \cite{bcnn05}, $D_s D^* K$, $D_s^* D K$ \cite{angelo06} and $D^* D^* \rho$ \cite{bmmf11}), motivated by the development of effective theories related to charm particles interaction, showed a similar conclusion in all of them.

 However, in recent years, due to the precise measurements of $B$ decays performed by BELLE, BES and BABAR, the physics of $B$ meson has gained a new relevance. The observed 
$B_{s1}(5830)$ by CDF Collaboration \cite{PRL100} and $B_{s2}(5840)$ by CDF and D0 Collaboration \cite{PRL101} stimulated our interest in the physics of bottom mesons interactions \cite{mm10}. In particular, the unobserved $ B^*_{s_0}(5725)$ state and the axial $B_{s_1}(5778)$ have been interpreted as hadronic bound states (hadronic molecules), because their masses are close to the thresholds of the corresponding hadronic pairs. In reference \cite{FGLM2008}, the authors used a phenomenological Lagrangian approach to calculate the strong and radiative decay of these new mesons, $ B^*_{s_0}(5725)$ and  $B_{s_1}(5778)$, 
which were considered as bound states of the $ B K $ and $ B^* K$  mesons respectively, to provide more information about some properties. In particular, the coupling constant of
$B^*_{s}B K$ vertex was an input in this calculation.
  
Considering that the form factors are introduced in the development of effective theories to study bottom interactions, in this paper, we calculated the form factors and coupling constant of the $ B^*_{s} B K $ vertex, using the QCDSR method, which is the only method that permits to extract the coupling constant of the three meson processes without any dependence on other empirical coupling constant. Therefore, we evaluate two form factors, one when $B$ is the off-shell particle and another when $K$ is the off-shell and extrapolate these results to obtain the coupling constant $g_{B_s^* B K}$. We also want to estimate the uncertainties of the QCDSR method by analyzing different sources of errors.  

  In section II, we describe the QCDSR technique; in II.A, the QCD side for this vertex 
and in II.B, the phenomenological side. In section III, we show the results; in section IV, 
we estimate the uncertainties and, finally,  conclude.

\section{The sum rule for the  $ B_s^* B K$  vertex}

Following our previous works, specifically in Ref.\cite{angelo06}, the three-point function associated 
with the $B_s^* B K$ vertex is givem by
\begin{equation}
\Gamma_{\nu \mu}^{(B)}(p,\pli)=\int d^4x \, d^4y \;\;
e^{i\pli\cdot x} \, e^{-i(\pli-p)\cdot y}
\langle 0|T\{j_{\nu}^{{K}} (x) j_{5}^{B \dagger}(y) j_{\mu}^{{B_s^{*}} \dagger}
(0)\}|0\rangle\, 
\label{correBoff} 
\end{equation}
for an off-shell $B$ meson, and:
\begin{equation}
\Gamma_{\mu}^{({K})}(p,\pli)=\int d^4x \, 
d^4y \;\; e^{i\pli\cdot x} \, e^{-i(\pli-p)\cdot y}\;
\langle 0|T\{j_{\mu}^{{B_s^*}}(x)  j_{5}^{{K} \dagger}(y) 
 j_{5}^{B \dagger}(0)\}|0\rangle\, ,
\label{correkoff} 
\end{equation}
for an off-shell ${K}$ meson. The expressions for the vertices 
(\ref{correBoff}) and (\ref{correkoff}) contain different numbers of 
Lorentz structures, and each structure can be a different Sum Rule (SR) 
to be considered.  
Equations~(\ref{correBoff}) and  (\ref{correkoff}) can be calculated in two 
different ways: using quark degrees of freedom --the theoretical or the QCD side--  
and using hadronic degrees of freedom --the phenomenological side. In the QCD side, the 
correlators are evaluated  using the Wilson operator product expansion (OPE),
which incorporates the effects of the QCD vacuum through an infinite 
series of condensates of in\-crea\-sing 
dimension. The phenomenological side is written in terms of 
hadronic degrees of freedom, which is the responsible for the introduction of the form 
factors, decay constants and masses. After performing a double Borel transformation, both 
representations are matched invoking the quark-hadron global duality.

\subsection{The OPE side}

In the OPE or theoretical side, each meson interpolating
field appearing in Eqs.~(\ref{correBoff}) and (\ref{correkoff}) can be written 
in terms of the quark field operators. For the $B$ off-shell case, Eq.~(\ref{correBoff}),
we use the following meson currents: 
\beqa
&&j^{B^*_s}_{\mu}(0)= \bar b \gamma_{\mu} s \nonumber \\
&&j^B_5(y)= i \bar q  \gamma_5 b  \\
&&{j^K_{\nu}}(x)=\bar q \gamma_{\nu}\gamma_5 s  \nonumber
\eeqa
and when the $K$ is off-shell, we have chosen the pseudoscalar current for it, following the same procedure as in previous works as $D^*D\pi$ \cite{nnb02} and $D^*_s D K $ \cite{angelo06}: 
\beqa
&&{j^K_{5}}(x)= i \bar q \gamma_5 s,
\eeqa 
this choice of current is motivated by the coupling constant that is obtained by extrapolating the $g^{(K)}_{B^*_s B K}(Q^2)$ to momentum $Q^2$ equal to the square mass of the particle, that in this case will be $m^2_K$. Otherwise, if we use the axial vector current when $K$ meson is off-shell, the sum rule obtained is also good and the coupling constant result is very similar to the pseudoscalar case.

The $u$, $d$, $s$ and $b$ are the up, down, strange and bottom quark fields, respectively. Each 
one of these currents has the same quantum numbers of the associated
mesons.

In order to obtain the QCD side, we use the Cutkosky rule and we obtain an 
expression where all the possible structures appear. 
We can obtain a sum rule for each structure that have an invariant amplitude, which 
can be written in a double dispersion relation over the virtualities $p^2$ and 
${\pli}^2$, holding $Q^2= -q^2$ fixed:
\begin{equation}
\Gamma^T(p^2,{\pli}^2,Q^2)=-\frac{1}{4 \pi^2}\int_{s_{min}}^\infty ds
\int_{u_{min}}^\infty du \:\frac{\rho^T(s,u,Q^2)}{(s-p^2)(u-{\pli}^2)}\;,
\;\;\;\;\;\; \label{dis}
\end{equation}
where $\rho^T(s,u,Q^2)$ equals the double discontinuity of the amplitude
$\Gamma^T(p^2,{\pli}^2,Q^2)$, $T$ index means $B$ off-shell or $K$ off-shell,
$q = \pli -p $ and $s_{min}$ and $u_{min}$ are integration limits equal to $s_{min}=(m_b +m_s)^2 $ and $u_{min}= t -m_b^2 + m_s^2$ for $B$ off-shell and  $s_{min}=m_b^2$ and $u_{min}= t + m_b^2$ for $K$ off-shell.

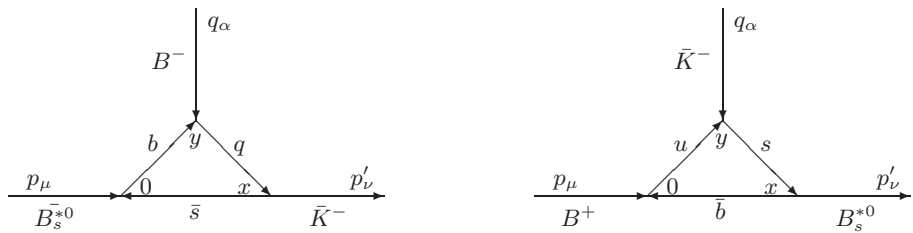
\begin{figure}[ht]
\begin{picture}(12,3.5)
% Left diagram
\put(0.0,0.5){\vector(1,0){1.5}}
\put(3.5,0.5){\vector(-1,0){2}}
\put(3.5,0.5){\vector(1,0){1.5}}
\put(1.5,0.5){\vector(1,1){1}}
\put(2.5,1.5){\vector(1,-1){1}}
%\put(2.5,1.5){\vector(0,1){1.5}}
\put(2.5,3){\vector(0,-1){1.5}}
\put(2.65,2.75){$q_{\alpha}$}
\put(0.25,0.65){$p_\mu$}
\put(4.55,0.65){$p'_\nu$}
\put(2.4,0.2){$\bar s$}
\put(1.85,1.1){$b$}
\put(3,1.1){$q$}
\put(2.4,1.2){$y$}
\put(1.75,0.53){$0$}
\put(3.05,0.53){$x$}
\put(1.9,2.2){$B^-$}
\put(0.35,0.1){$\bar{B_s^{*0}}$}
\put(4,0.1){$\bar{K}^-$}
% Right diagram
\put(7,0.5){\vector(1,0){1.5}}
\put(10.5,0.5){\vector(-1,0){2}}
\put(10.5,0.5){\vector(1,0){1.5}}
\put(8.5,0.5){\vector(1,1){1}}
\put(9.5,1.5){\vector(1,-1){1}}
\put(9.5,3){\vector(0,-1){1.5}}
\put(9.65,2.75){$q_\alpha$}
\put(7.25,0.65){$p_\mu$}
\put(11.55,0.65){$p'_\nu$}
\put(9.4,0.2){$\bar b$}
\put(8.85,1.1){$u$}
\put(10,1.1){$s$}
\put(9.4,1.2){$y$}
\put(8.75,0.53){$0$}
\put(10.05,0.53){$x$}
\put(8.85,2.2){$\bar{K}^-$}
\put(7.35,0.1){$B^+$}
\put(11,0.1){$B_s^{*0}$}
\end{picture}
\caption{Perturbative diagrams for the $B$ off-shell (left) and $K$
off-shell (right) correlators. }
\label{fig1}
\end{figure}

The invariant amplitudes receive contributions 
from all terms in the OPE. The first of those contributions comes from 
the perturbative term, which is represented in Fig.~\ref{fig1}.

In order to obtain the form factor, we choose one of the structures in Eqs.(\ref{correBoff}) and (\ref{correkoff}), the one that has less ambiguities in the QCD sum rule approach, which means less influence from the higher dimension condensates, better stability as a function of the Borel mass and has larger pole than continuum contribution. 
The other structures that are ``good "  sum rules can also be considered to improve the uncertainties of the method.
For $B$ off-shell meson, which has two different structures, we choose 
the $p'_{\mu} p'_{\nu}$ because it is the best structure with less 
ambiguities.  
For $K$ off-shell meson, we have $p_{\nu}$ and $p'_{\nu}$ structures, 
both being excellent sum rules. The chosen is $p'_{\nu}$ and the other will be used in the calculation of the uncertainties. 

 The corresponding perturbative spectral densities, which enter in Eq.~(\ref{dis}), are: 
\begin{equation}
\rho^{(B)}(s,u,Q^2)=\frac{3 }{2\pi\sqrt\lambda}
\left[\left(2 m_b (A-B) \right) \right] \,
\label{rhoBoff}
\end{equation}
for the $p'_{\mu} p'_{\nu}$ structure of the $B$ off-shell case, and 
\begin{equation}
\rho^{(K)}(s,u,Q^2)=-\frac{3}{2\pi\sqrt\lambda} 
\left[B \left(\frac{u-s-t}{2} - m_s^2 +m_b^2 +m_b m_s\right) +\frac{s+m_s^2-2m_b^2}{2}\right] 
\label{rhokoff}
\end{equation}
for the $p'_{\nu}$ structure of the $K$ off-shell case, where $\lambda = \lambda(s,u,t) = 
s^2+t^2+u^2-2st-2su-2tu$, $s=p^2$, $u=p'^2$, $t=-Q^2$ 
and $A$ and $B$  are functions of $(s,u,t)$, givem by 
the following expressions:
\begin{eqnarray}
A=-\frac{\pi\overline{|\vec k|}^2 }{\overline{|\vec p'|}^2} 
\left(3 \cos\overline{\theta} -1 \right) \label{D}; \;\;\;\;\;\;\;\;\;\;\;
B= 2\pi\frac{\overline{|\vec k|}}{ \overline{|\vec p'|}} \cos\overline{\theta} \,\, 
\label{N} ; 
\end{eqnarray}
where 
\begin{eqnarray}
\overline{|\vec k|}^2&=& \overline{k_0}^2-m_i^2   \label{vk}; \;\;\;\;\;\;
\cos\overline{\theta}=-\frac{2p'_0\overline{k_0}-u-m_i^2 -\eta m_b^2}
{2 \overline{|\vec p'|} \overline{|\vec k|}}   \label{ctheta};\nonumber \\
p'_0&=&\frac{s+u-t}{2\sqrt{s}}    \label{pl0}; \;\;\;\;\;\;\; 
\overline{|\vec p'|}^2=\frac{\lambda}{4s}     \label{vpl};\;\;\;\;\;\;\
\overline{k_0}= \frac{s+ m^2_i-\epsilon \; m_b^2}{2\sqrt{s}}   \label{k0b} ;
\end{eqnarray}
with $i= s $, $\eta=0$ and $\epsilon=1$ for $B$ off-shell and $i= b $, $\eta=-1$ and $\epsilon=0$ for $K$ off-shell.

The main OPE contribution is the perturbative. The next terms in the OPE expansion are the quark condensate, the gluon condensate and the mixed quark-gluon condensate.
In this work, the quark condensate was calculated and has the following expression obtained for the $K$ off-shell case:
\begin{equation}
\Gamma_{\nu}^{< \bar b  b>}= m_s < \bar b\; b> e^{m_s^2/M'^2}  p_{\nu}
\end{equation}
this contribution is proportional to the bottom condensate, consequently, the contribution is null \cite{SVZ}. The other two quark condensates contributions for the $K$ off-shell case vanish after the application of the double Borel transformation.
For $B$ off-shell case, the quark condensate does not contribute for the structure considered.
In this calculation, we do not include gluon condensate and mixed quark gluon condensate, because as seen in previous calculation of the form factors of $D^*D\pi$ and $B^*B \pi$ vertices \cite{nnb02}, we found out that the gluon contribution is negligible as compared with the perturbative one. 
Moreover, the gluon condensate decreases with the Borel mass and the most important role of the gluon condensate is to guarantee the stability as a function of the Borel mass. In this vertex,
this stability is guaranteed together with a good contribution of the pole. 

\subsection{The phenomenological side}

 The three-point functions from Eqs.~(\ref{correBoff}) and 
(\ref{correkoff}), when written in terms of hadron masses, decay constants and form factors, are the phenomenological side of the sum rule, which is based on the interactions at the hadronic level. These interactions are described here by the following effective Lagrangian  \cite{lagrangiana}:
\beq
\mathcal{L}_{B_s^* B K}=
ig_{B_s^* B K} \Big [  B_s^{*\mu} ( \bar{B} \partial_{\mu}K  
- \partial_{\mu}\bar{B} K )                   
-  B_s^{*\mu} ( \bar{K} \partial_{\mu}B 
- \partial_{\mu} B \bar{K}  )    \Big ]\;,
\nonumber
\label{lagr}
\eeq
from where we can extract the matrix element associated with the
$B_s^* B K$ vertex.

The meson decay constants $f_{K}$, $f_B$ and $f_{B_s^*}$ are
defined by the following  matrix elements:
\beq
\langle 0|j_{\nu}^{K}|{K(p)}\rangle= i  f_{K} p_{\nu},
\label{fK}
\eeq
\beq
\langle 0|j_{\mu}^{B_s^{*}}|{B_s^{*}(p)}\rangle= m_{B_s^{*}} f_{B_s^{*}} 
\epsilon_{\mu}^*(p) \, ,
\label{fBss}
\eeq
and 
\beq
\langle 0|j_{5}^{B}|{B(p)}\rangle= \frac{m_{B}^2}{m_b} f_{B}  \, ,
\label{fB}
\eeq
where $\epsilon_{\mu}^*$ is the polarization vector  for the $B_s^{*}$ meson. 
Saturating Eqs.~(\ref{correBoff}) and (\ref{correkoff}) with 
$K$ and $B^*_s$ states and using Eqs.~(\ref{fK}) - (\ref{fBss}), we arrive at
\beqa
&&\Gamma_{\mu \nu }^{(B)}(p,p^\prime,q)=  g^{(B)}_{B_s^* B K}(q^2)
\frac{f_{B_s^*}f_{K}f_{B}m_{B_s^*}\frac{m^2_{B}}{m_b}}
{(p^2-m^2_{B_s^*})(q^2-m^2_{m_B})({p^\prime}^2 -m^2_{K})} \nonumber \\
& & \times  \left[2 \pli{_\mu} \pli_{\nu}   
 - \pli_{\mu}p_{\nu} \left(1 +\frac{(m_K^2-q^2)}{m_{B^*_s}^2}\right) \right]+ 
{\textit{``continuum''}},
\label{phenBoff}
\eeqa
when $B$ is off-shell. Using the matrix element of $K$ meson equal to
\beq
\langle 0|j_{5}^{K}|{K(p)}\rangle= \frac{m_{K}^2}{m_s} f_{K}  \, ,
\label{fK2}
\eeq
we arrive at an expression for $K$ off-shell: 
\beqa &\Gamma^{(K)}_{\mu }(p,p^\prime,q)&=- g^{(K)}_{B^*_s BK}(q^2)\frac{f_B  f_K f_{B^*_s}
m_{B^*_s}^2 m_B^2 m_K^2}{m_b m_s(p^2-m_B^2) ({p^\prime}^2-m_{B_s^*}^2) (q^2-m^2_K)}\nonumber \\
&&\times \left[ 2 p_{\mu}
+\pli_{\mu} \left(1+\frac{(m_B^2-q^2)}{m_{B^*_s}^2}\right)\right]+{\textit{``continuum''}}.
\label{phenKoff} 
\eeqa
In the Eqs.~(\ref{phenBoff}) and (\ref{phenKoff}), we can see all the structures that appear in the QCDSR and the {\it``continuum"} contribution, which may be subtracted assuming that its contribution is equal to the contribution of the QCD side, by introducing the continuum thresholds $s_0$ and $u_0$ \cite{io2} as upper integration limits in Eq.~(\ref{dis}).

In order to improve the matching between the two sides of the sum rule,
we perform a double Borel transformation \cite{io2} in the variables 
$P^2=-p^2\rightarrow M^2$ and $P'^2=-{\pli}^2\rightarrow M'^2$, 
on both invariant amplitudes $\Gamma^{(T)}$ (QCD side) and $\Gamma^{(T)}_{ph}$ (phenomenological side). 
Equating the results, we obtain the form factors 
$g^{(T)}_{B^*_s B K }(Q^2)$ that appear in 
Eqs.~(\ref{phenBoff}) and (\ref{phenKoff}).

\section{Results and discussion}

 Table \ref{param} shows the parameters used in the present 
calculation. We have used the experimental value for $f_{K}$ of Ref.~\cite{fkvalue}, for $f_{B^*_s}$  and $f_{B}$ from Ref.~\cite{fbsvalue}; 
for $m_s$ from Ref.~\cite{msvalue} and for $m_b$ we used recent results
of Ref.~\cite{mbvalue,narisson}.
The continuum thresholds are givem by $s_0=(m_i+ \Delta_s)^2$ and $u_0=(m_{o}+\Delta_u)^2$, where $m_i$ is the mass of the incoming meson, $m_{o}$ is the mass of the out coming meson and $ \Delta_u $ and $ \Delta_s$ are usually around $0.5 \; GeV$, but this value can be varied to obtain a good pole-continuum contribution and stability of the sum rule with the Borel mass parameter. 
\begin{table}[ht]
\begin{center}
\begin{tabular}{|c|c|c|c|c|c|c|}\hline
   & $q$ & $s$ & $b$ & $K$ & $B$ & $B^*_s$\\\hline
  $m$ (GeV) & $0.007$ & $0.13$ & $4.20$ & $0.49$ & $5.27$ & $5.41$\\\hline
  $f$ (MeV)&-&-&-&$160$ &$208$ &$250$ \\\hline
\end{tabular}
\caption{Parameters used.}\label{param}
\end{center}
\end{table}

%------------------ For B meson off shell 
\subsection{ B off-shell form factor}

In order to obtain the $B$ off-shell form factor, we choose the structure $p'_{\nu}p'_{\mu}$ which has an excellent stability and its pole contribution is larger than the continuum contribution in a Borel region between $10 < M^2 < 24 \; GeV^2 $. 
 The first order in the OPE, the quark condensate does not contribute in this structure. 
In this paper, the next order in the OPE, the gluon condensate was not calculated, because in our previous work of $D^* D \pi$ form factor \cite{nnb02} and
also in $J/\Psi D^* D$ \cite{gluonofmarina}, we found out that the gluon condensate contribution is less than 10\% for a Borel mass of $ 5 \; GeV^2$.
 In this case, we are dealing with heavy mesons and so we are working with a Borel mass that is $21 \; GeV^2$ with which we are expecting the gluon condensate to be
negligible and it does not contributing to modify the result of the perturbative term.

\begin{figure}[ht] 
%\begin{center}
{\epsfig{height=55mm,figure=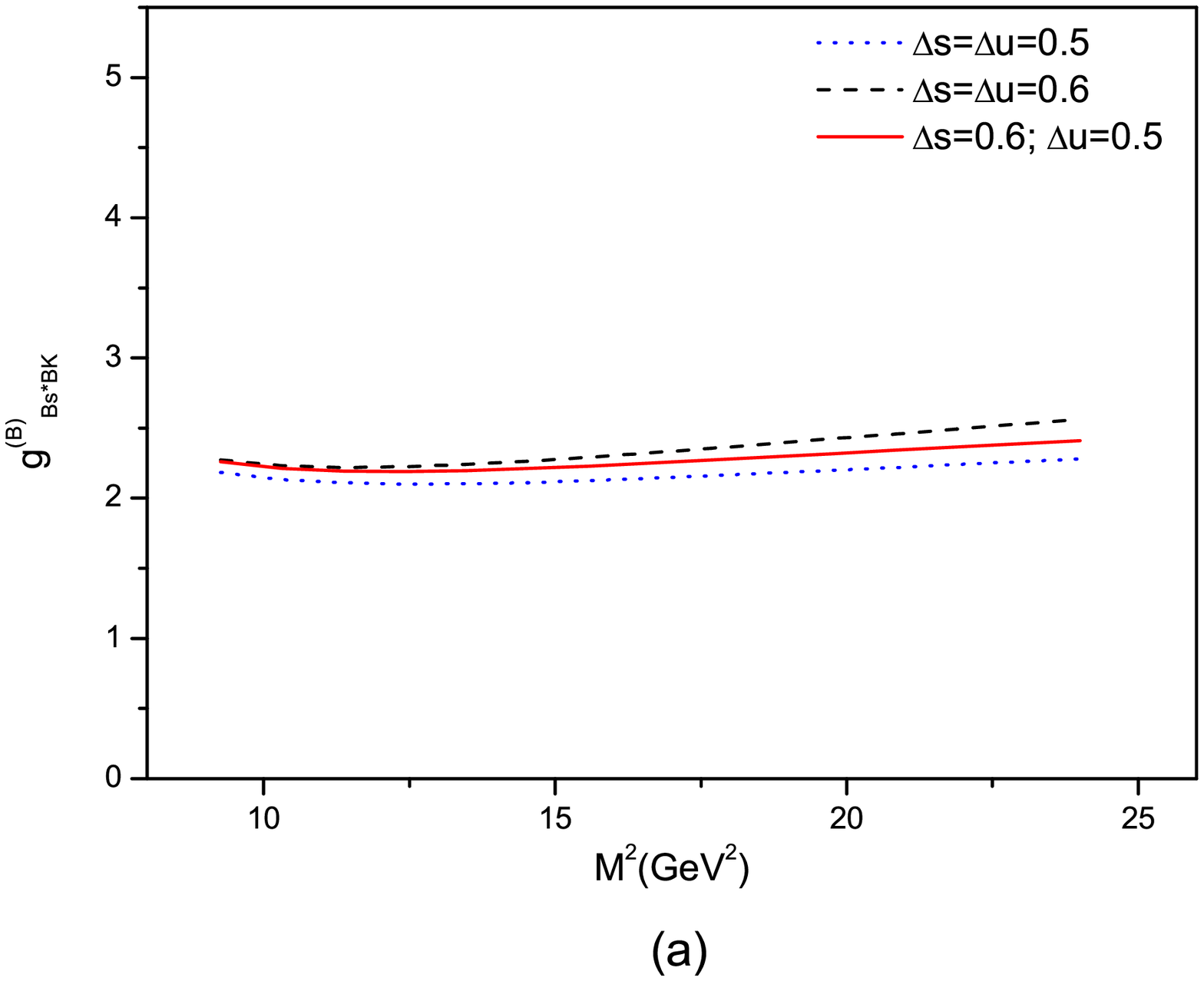}}
{\epsfig{height=55mm,figure=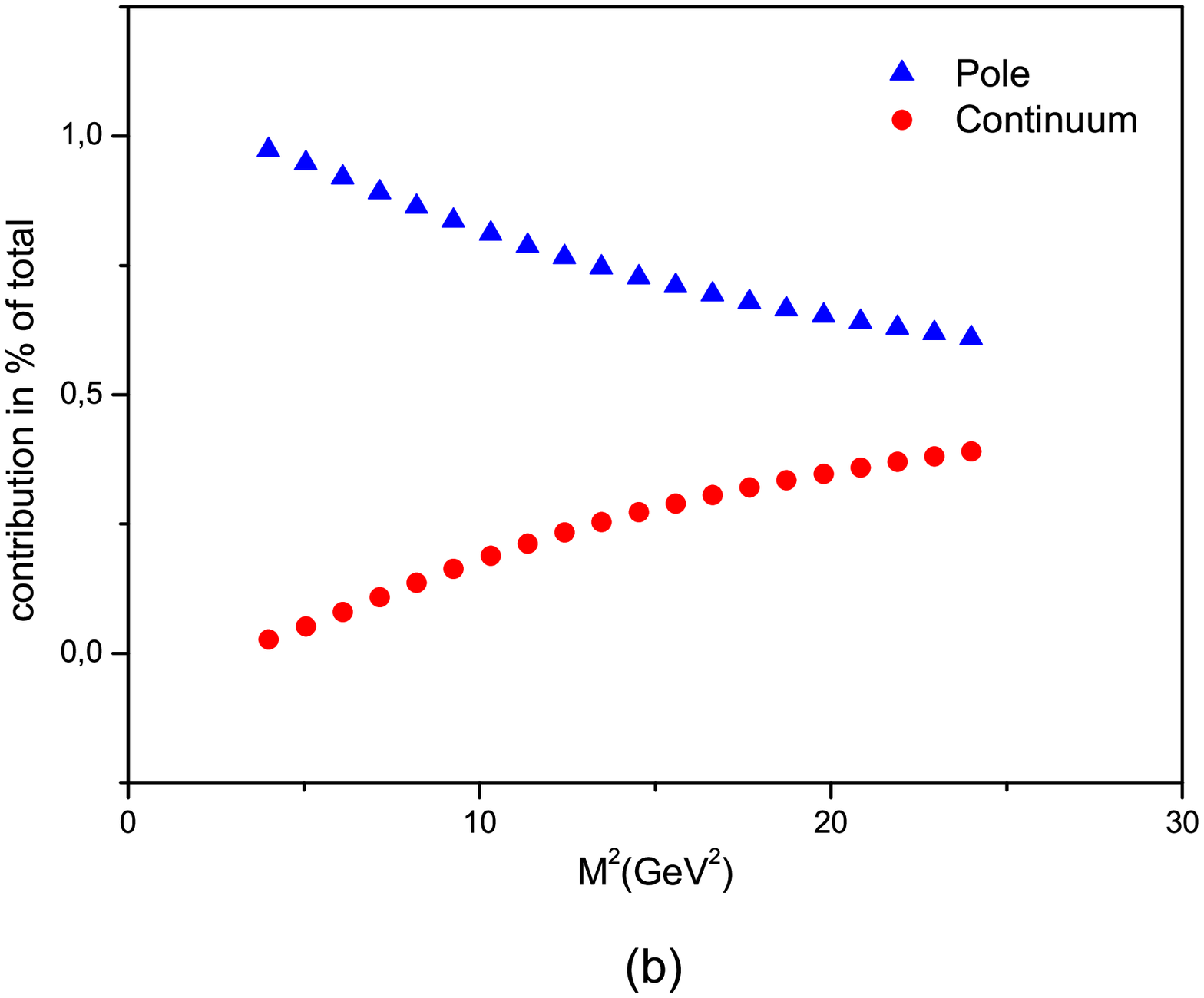}}
\caption{a) $g^{(B)}_{B^*_s B K}(Q^2=1 \GeV^2)$ as a function 
of the Borel mass $M^2$ and different thresholds and b) pole-continuum contributions.}
\label{estabilidadeBoff}
%\end{center}
\end{figure}

 Performing the match beetwen the QCD and the phenomenological side,  in Fig.~(\ref{estabilidadeBoff}a), we show the Borel window stability for different value's combinations of thresholds and in the same Fig.~(\ref{estabilidadeBoff}b), we can see the pole and continuum contributions observing that the pole contribution is always bigger than the continuum contribution in the same Borel window of stability, therefore we have good credibility for this sum rule. Joining this criteria with the thresholds $\Delta_s= 0.6 \GeV$ and $\Delta_u = 0.5 \GeV $, we obtain the form factor for $B$ off-shell: $g^{(B)}_{B^*_s B K}(Q^2)$, where we have used  $\frac{M^2}{M'^2} = \frac{m^2_{K}}{m^2_{B^{*}_s} -\;{m^2_{ b}} } $  as a Borel mass relation, which is usual when light and heavy mesons are involved.

%-----ESta figura saiu para estar junto a estabilidade ---------
%\begin{figure}[h!] 
%%\begin{center}
%\centerline{\epsfig{figure=polo-cont-B-off.eps,height=80mm}}
%\caption{Pole (triangles) and continuum (circles) contribuition to  
%$g^{(B)}_{B^*_s B K}(Q^2=1\GeV^2, M^2)$, 
%as a function of the Borel mass $M^2$.}
%\label{pcBoff}
%%\end{center}
%\end{figure} 

%----------FORM FACTOR b off  - Tiramos esta para o paper final----------
%\begin{figure}[b!] 
%\centerline{\epsfig{figure=form-factor-B-off.eps,height=80mm}}
%\caption{$g^{(B)}_{B^*_s B K}$ (triangles) QCDSR form factor as a function of
%$Q^2$. The solid line correspond to the monopolar parametrization (Eq.~\ref{monoBoff} of the QCDSR %results.}
%\label{formfactorBoff}
%\end{center}
%\end{figure}

The triangles in Fig.~(\ref{erros}) correspond to the sum rule result, which can be 
fitted by the monopolar function represented by the dashed line in 
Fig.~(\ref{erros}) and given by  

%(shown by the solid line in Fig.~\ref{formfactorBoff}):
\begin{equation}
g_{B^*_s B K}^{(B)}(Q^2)= \frac{84.42}{ 34.99 +Q^2} 
\label{monoBoff}
 \end{equation}
The coupling constant is obtained when the form factor is
extrapolated to $Q^2= -m^2_{T}$, where $m_{T}$ is the mass of the off-shell meson \cite{bclnn01,angelo06,bmmf11}. 
Using $Q^2=-m_{B}^2$ in Eq~(\ref{monoBoff}), the resulting coupling 
constant is equal to: 
\begin{equation}
g_{B^*_s B K}= 10.6 
\label{couplingboff}
\end{equation}

For the $B$ meson off-shell case, we have other structure, $p'_{\nu}p_{\mu}$, to work with. This sum rule has quark condensate contribution but it has not good stability and its continuum contribution is larger than the pole contribution, therefore this sum rule is not considered a good sum rule.

\subsection{ $K$ off-shell form factor} 

\begin{figure}[ht] 
%\begin{center}
{\epsfig{figure=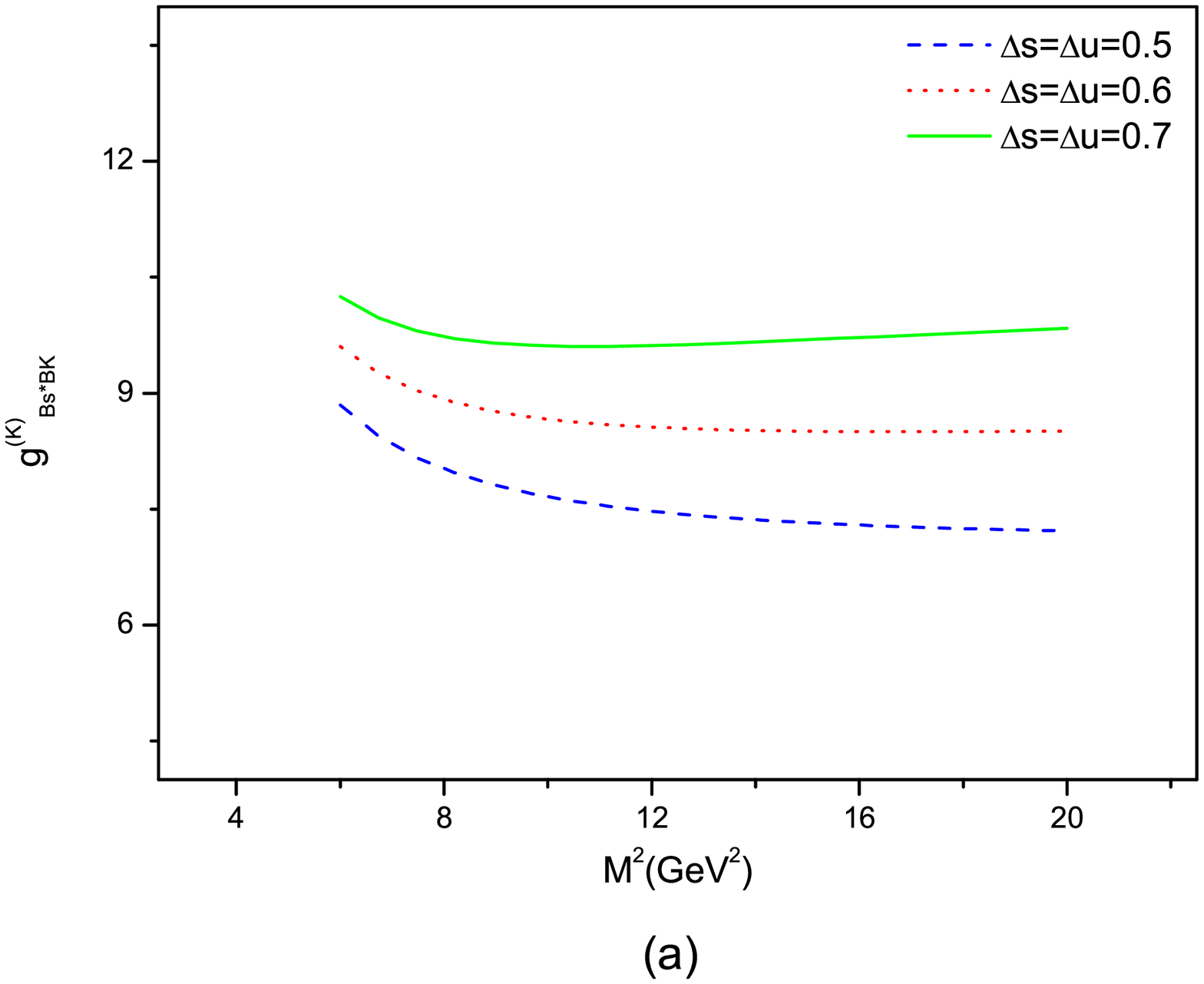,height=55mm}}
{\epsfig{figure=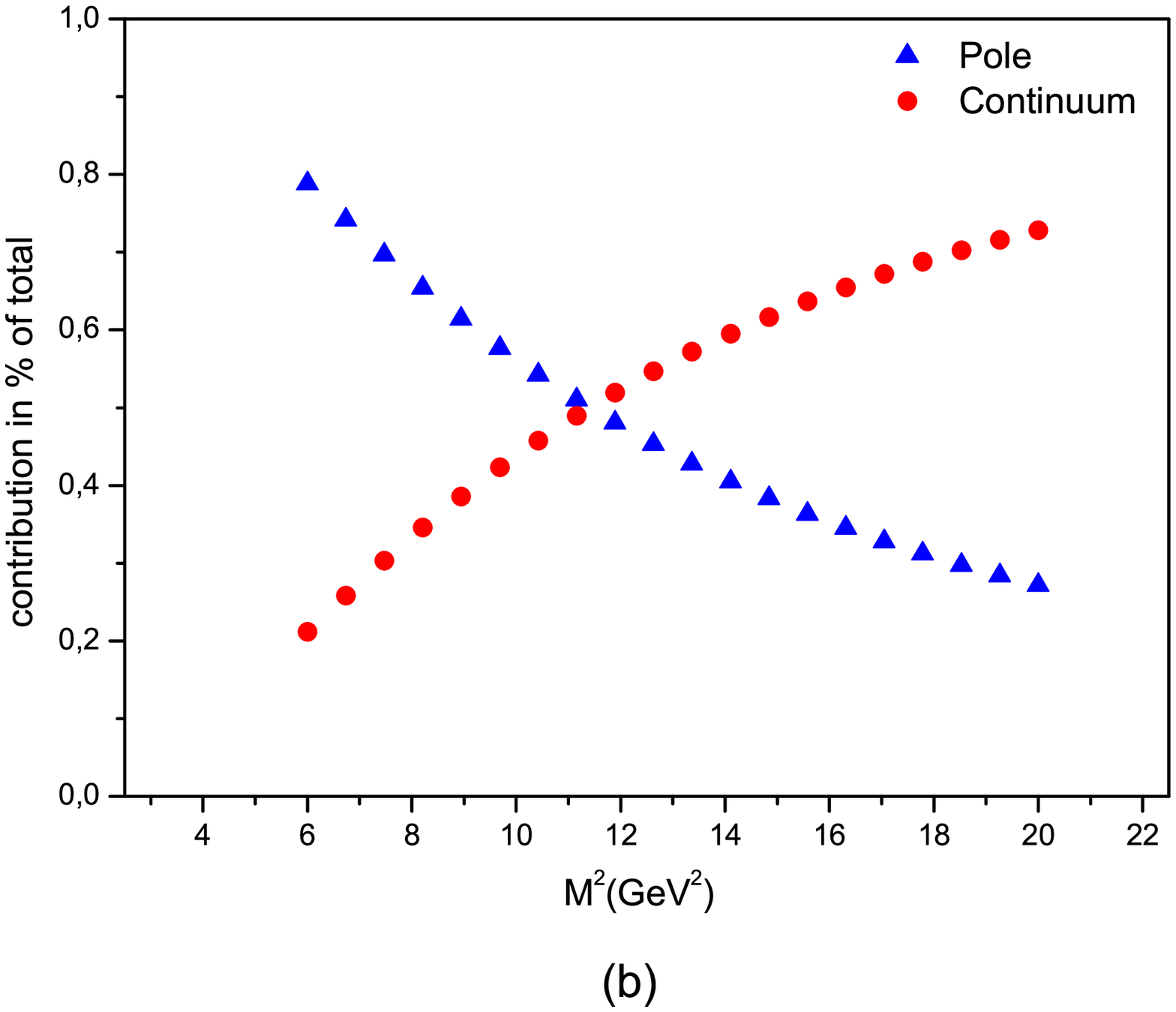,height=55mm}}
\caption{a) $g^{(K)}_{B^*_s BK}(Q^2 = 2 \; GeV^2)$ stability as a
function of the Borel mass for different
threshold values, structure $ p^\prime_{\nu}$ and b) pole and continuum contributions. }
\label{estabilidadeKoffpl}
%\end{center}
\end{figure}

In the case $K$ off-shell, we have two structures in Eq.~(\ref{phenKoff}) that can be used:  $p'_{\nu}$ and $ p_{\nu}$.  Both structures give good sum rule results, that means a good pole-continuum contribution and good stability. The quark condensate of the bottom is
considered, but the contribution is null.
At first, we have chosen the $p'_{\nu}$ structure to obtain the form factor. The stability for different thresholds are showed in Fig.~(\ref{estabilidadeKoffpl}a).  
For $\Delta_s= 0.7 \GeV$, $\Delta_u = 0.7 \GeV $ and for a Borel mass of $ 7.2 \GeV ^2 $, we obtain a pole contribution of about 70\% of the total contribution ( Fig.~(\ref{estabilidadeKoffpl}) part b)).

%------k  off P' --------------- 

%\begin{figure}[ht] 
%\begin{center}
%\centerline{\epsfig{figure=polo-cont-K-pl.eps,height=80mm}}
%\caption{Pole versus continuum contributions  to  $g^{(K)}_{B^*_s %BK}(Q^2=1\GeV^2)$  
%as a function of the Borel mass $M^2$, in the $p'$ structure.}
%\label{pckoffpl}
%\end{center}
%\end{figure}

The white circles in Fig.~(\ref{erros}) represent the sum rule result fitted by 
a Gaussian function that is the short dashed line in the same figure and is given by
\begin{equation}
g_{B^*_s B K}^{(K)}(Q^2)= 10.71 e^{- (Q^2/7.49)^2} \;,
\label{gaussKoff}
\end{equation}
by extrapolating Eq.~(\ref{gaussKoff}) to $Q^2 = - m^2_K$, we obtain the coupling constant:
\begin{equation}
g_{B^*_s B K}= 10.7
\label{couplingkoffpl}
\end{equation}
% --------------------- Figura do fator de forma K off pl -saiu
%\begin{figure}[ht!] 
%\centerline{\epsfig{figure=form-factor-K-off-pl.eps,height=80mm}}
%\caption{$g^{(K)}_{B^*_s BK}$ (triangles) QCDSR form factors as a function of
%$Q^2$. The solid line correspond to the gaussian parametrization of the %QCDSR results.}
%\label{formfactorkoffpl}
%\end{center}
%\end{figure}
We can see an excellent agreement between the results obtained in Eqs. 
(\ref{couplingboff})  and (\ref{couplingkoffpl}), as expected.

We noted the same vertex behavior concerning the shape of the form factor that we obtained in previous works, when the heavy meson is the off-shell 
particle, the form factor is ``harder'' and has a big cutoff parameter $\Lambda$, which is defined from Eqs. (\ref{monoBoff}) and (\ref{gaussKoff}),
which have a general Gaussian form:
\begin{equation}
g_{B^*_s B K}(Q^2)= A \; e^{- (Q^2/{\Lambda})^2} 
\end{equation}
and a  monopolar form 
\begin{equation}
g_{B^*_s B K}(Q^2)= \frac{A}{ {\Lambda} + Q^2} 
\end{equation}
where $\Lambda$ parameter is equal to $ 34.99 \; GeV^2$ for $B$ off-shell meson and
is $7.49 \; GeV^2$ for $K$ off-shell meson.

%\begin{figure}[h!] 
%\centerline{\epsfig{figure=form-factor-B-Kpl.eps,height=80mm}}
%\caption{$g_{B^*_s BK}(Q^2)$ with B off-shell and K off-shell. }
%\label{bkplformfactor}
%\end{center}
%\end{figure}

\section{Uncertainties of the QCDSR method} 

The uncertainties in the QCDSR calculation come from different sources such
as decay constant errors, mass uncertainties, other ``good " structures which also can be considered and the QCDSR variation parameters themselves. In the next section, we discuss each case separately in sub-sections and use the results to find an error of the method in our result.

\subsection{$B_s^*$ off-shell}

\begin{figure}[ht] 
%\begin{center}
{\epsfig{figure=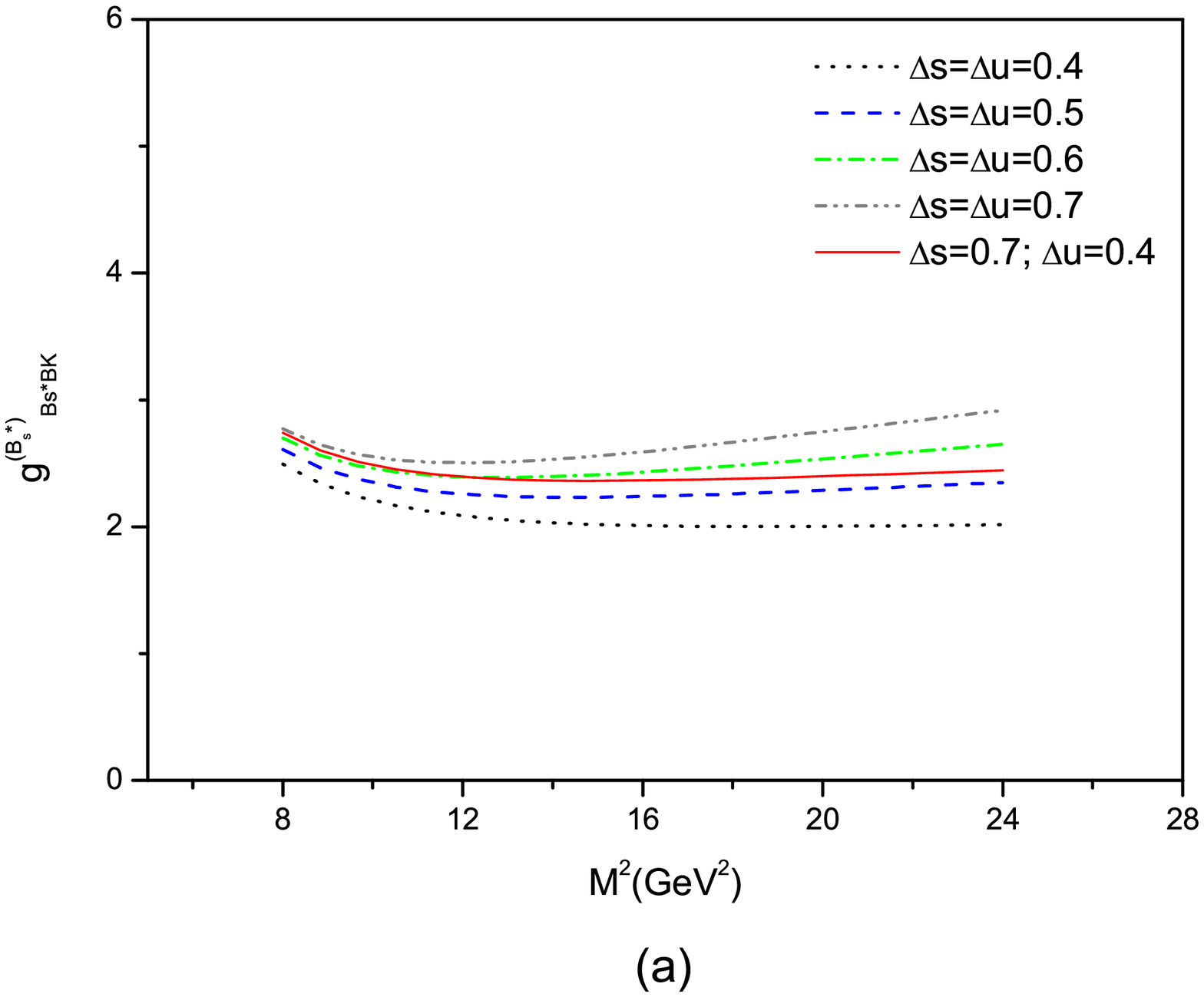,height=55mm}}
\epsfig{figure=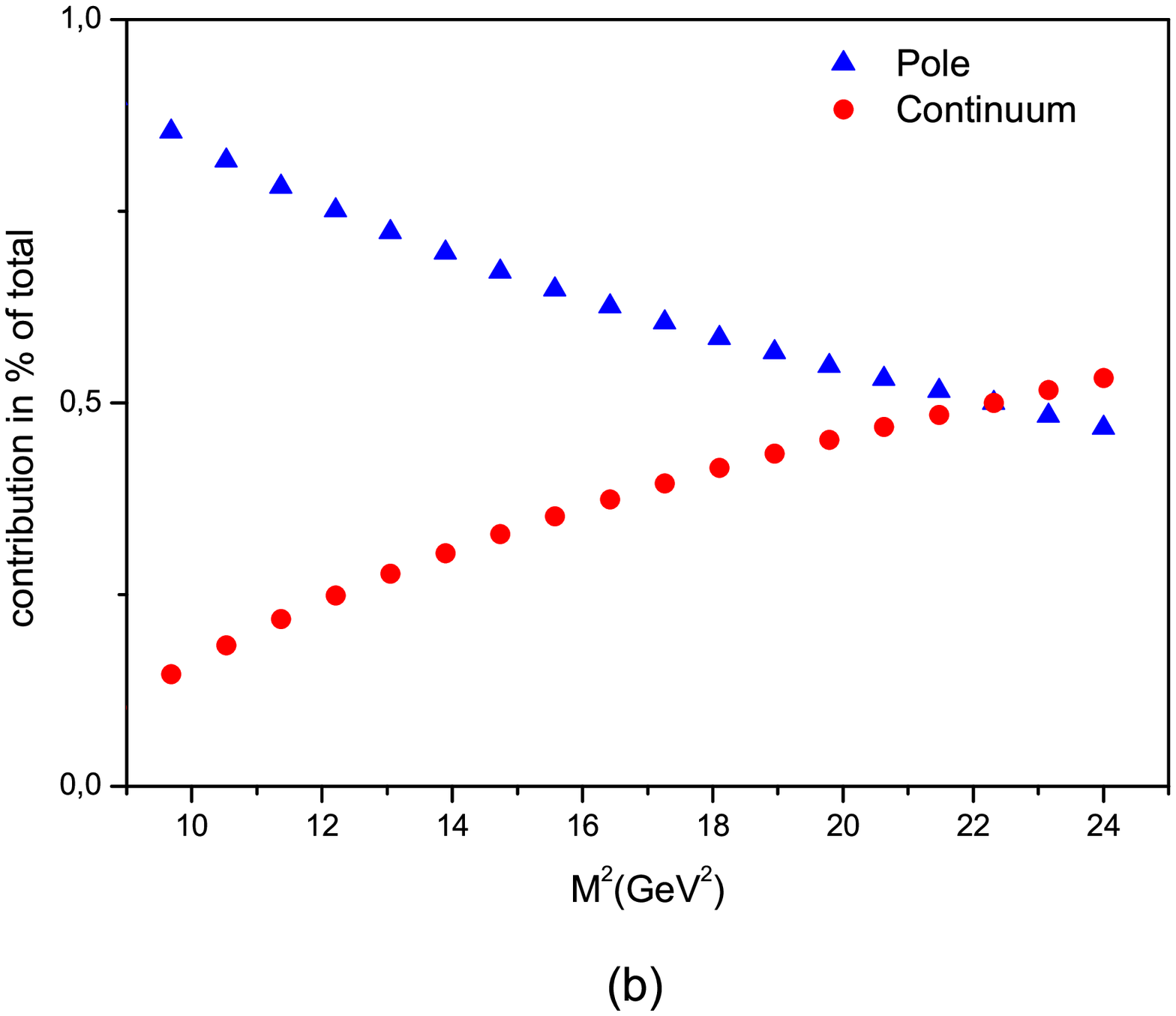,height=55mm}
\caption{(a) Stability of $g^{(B^*_s)}_{B^*_s BK}(Q^2=1 \; GeV^2)$ as
Borel mass and for different thresholds and (b) pole-continuum contributions.}
\label{estabilidadeBss}
%\end{center}
\end{figure}

At first, we calculate a third sum rule with the other heavy
meson of the vertex off-shell: the $B_s^*$. Since this meson has a small mass difference when compared with the $B$ meson mass, we expect to obtain a similar result to the one obtained for $B$ off-shell.
 We have two structures, $\pli_{\mu}\pli_{\nu}$ and $p_{\mu} \pli_{\nu}$, to work with in the method.  
The $p_{\mu}\pli_{\nu}$ structure does not present good stability in the Borel mass, therefore it is not considered.
The other structure, $\pli_{\mu}\pli_{\nu}$, has good stability and the pole is larger than the continuum contribution, for $ \Delta_s = 0.7 \; GeV $ and $\Delta_u =0.4 \;GeV$. 
Fig.~\ref{estabilidadeBss} a)  shows the Borel mass stability and  in b) the pole-continuum contribution. 

% --------------B*s off-pp'

%\begin{figure}[ht] 
%\begin{center}
%\centerline{\epsfig{figure=polo-cont-Bs-off.eps,height=80mm}}
%\caption{Pole versus continuum contributions  to  $g^{(B^*_s)}_{B^*_s %BK}(Q^2=1\GeV^2)$  
%as a function of the Borel mass $M^2$, in the $p_{\mu} \pli_{\nu}$ structure.}
%\label{pcBss}
%\end{center}
%\end{figure}

We extrapolate the QCDSR results, the dots in Fig.~\ref{erros},  by using a monopolar function until reaching $Q^2 = - m^2_{B^*_s}$,  and we obtain 
\begin{equation}
g_{B^*_s B K}^{(B^*_s)}(Q^2)= \frac{93.03}{37.80 + Q^2} \; 
\label{monoBsoff}
\end{equation}
for the form factor, and 
\begin{equation}
g_{B^*_s B K}= 10.25
\label{couplingbssoff}
\end{equation}
for the coupling constant.
% ---------------------form factor Bs off 
%\begin{figure}[b] 
%\centerline{\epsfig{figure=form-factor-Bs.eps,height=50mm}}
%\caption{$g^{(B^*_s)}_{B^*_s BK}$ form factor, structure $p_{\mu}\pli_{\nu}$ }
%\label{form-factor-Bs}
%\end{center}
%\end{figure}
Both results, Eq.~(\ref{monoBsoff}) and Eq.~(\ref{monoBoff}) show the same extrapolation function with similar parameters. 

%\sout{Using the three preceding sum rules, which were obtained for $K$, $B$, $B^*_s$ off-shell %cases, we find the mean value coupling constant }
%\begin{equation}
%\sout{\bar {g}_{B^*_s BK}= 10.52 \pm 0.03}
%\end{equation}
%\sout{where the error of $\pm 0.03$  was obtained without any \sout{parameter} \bruno{parameters'} %variations. }

\subsection{$K$ off-shell, structure $p_{\nu}$}

\begin{figure}[ht] 
%\begin{center}
{\epsfig{figure=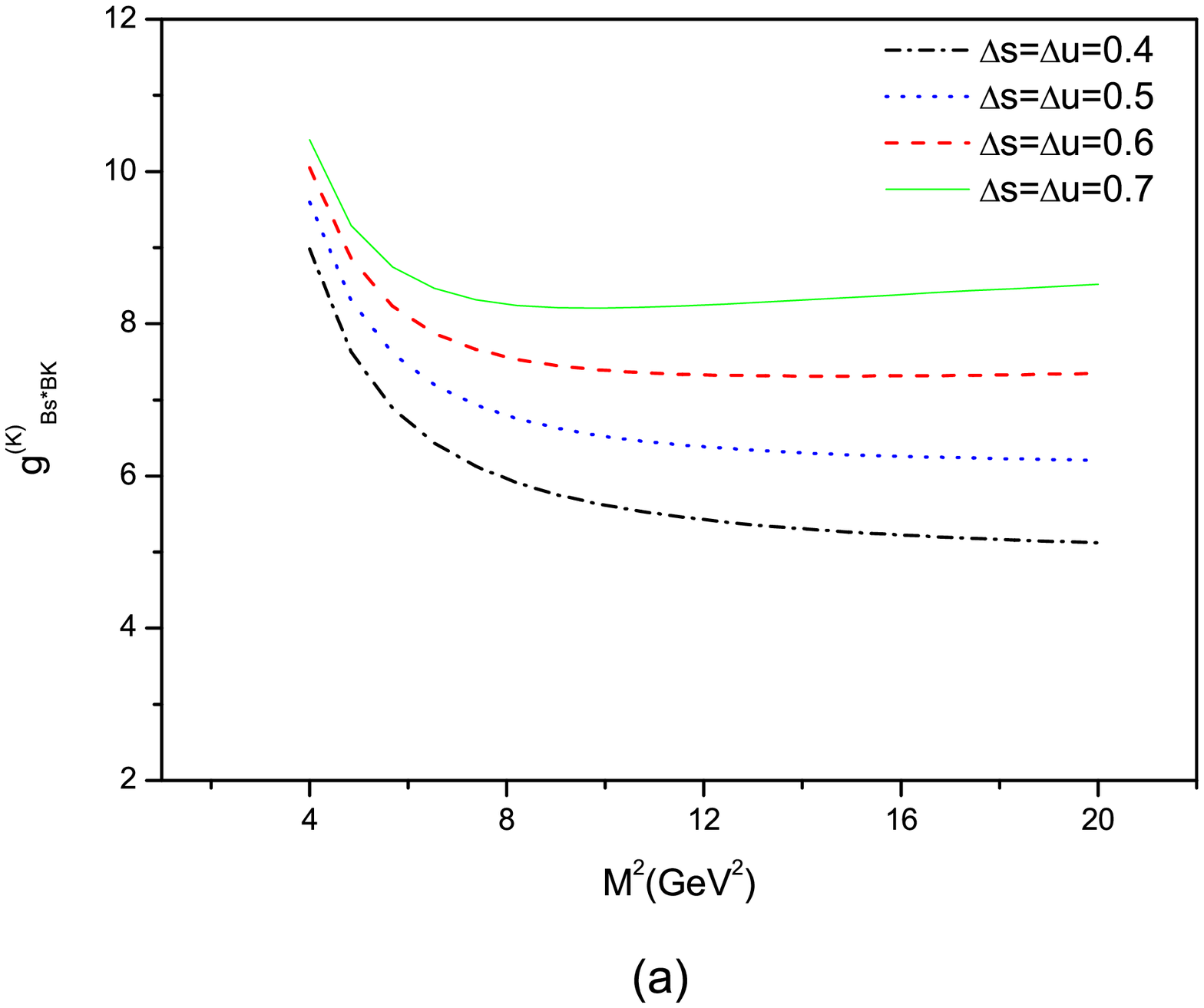,height=55mm}}
\epsfig{figure=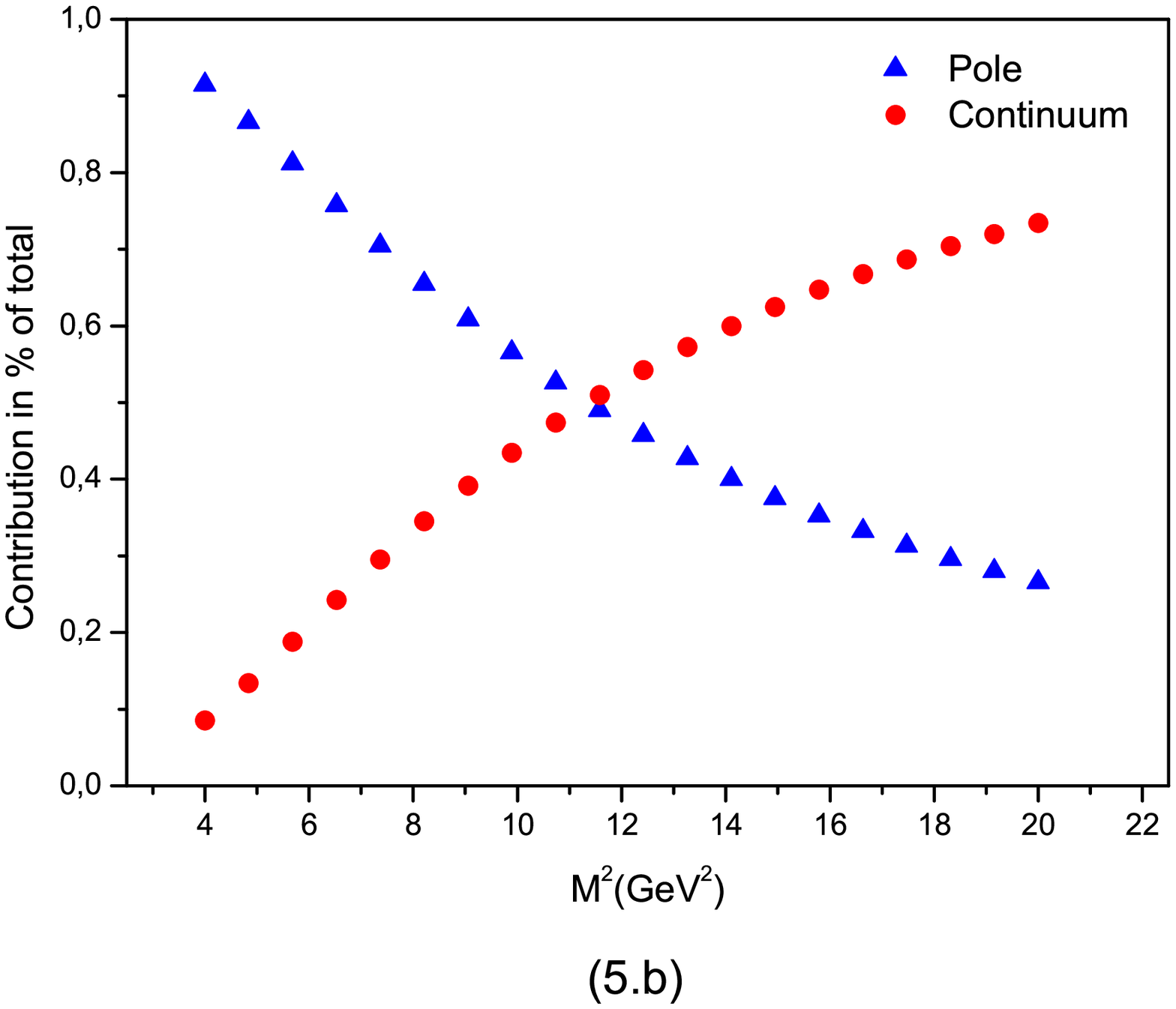,height=55mm}
\caption{a) Stability of $g^{(K)}_{B^*_s BK} (Q^2=2\;GeV^2)$ as a
function of the Borel mass, for different threshold
values and structure $p_{\nu}$; b) Pole and continuum contributions.}
\label{estabilidadeKoffp}
%\end{center}
\end{figure}

For the light meson off-shell, $K$, we also have the $p_{\nu}$ structure, which gives a good sum rule and is used here to calculate the uncertainties of the method. 
For this structure, we analyze the Borel mass stability for different threshold parameters, Fig.~\ref{estabilidadeKoffp} a), and the pole-continuum contribution, showed in Fig.~\ref{estabilidadeKoffp} b). 
% --------------K off-p
%\begin{figure}[ht] 
%\begin{center}
%\centerline{\epsfig{figure=polo-cont-K-p.eps,height=80mm}}
%\caption{Pole versus continuum contributions  to  $g^{(K)}_{B^*_s BK}(Q^2=1\GeV^2)$  
%as a function of the Borel mass $M^2$, in the $p_{\mu}$ structure.}
%\label{pcKoffp}
%\end{center}
%\end{figure}
We obtain a Gaussian function by extrapolating the QCDSR result given by: 
\begin{equation}
g_{B^*_s B K}^{(K)}(Q^2)= 10.65 e^{- (Q^2/4.99)^2} \;,
\label{gaussKoffpl}
\end{equation}
and making $Q^2 = - m^2_K$, we obtain the coupling constant of the vertex which is equal to: 
\begin{equation}
g_{B^*_s B K}= 10.7
\label{couplingkoffp}
\end{equation}
In Fig.~(\ref{erros}) we summarize the results obtained
with both structures showing an excellent agreement.  
% ---------------------form factor K off pl -SAIU 
%\begin{figure}[h!] 
%\centerline{\epsfig{figure=form-factor-K-off-p.eps,height=80mm}}
%\caption{$g^{(K)}_{B^*_s BK}$ form factor, structure $p_{\mu}$ }
%\label{formfactorkp}
%%\end{center}
%\end{figure}

\subsection{Uncertainties of parameter variations.}

In order to know, how much does the QCDSR depend on the parameters' variations,  we analyze, in this work, the experimental data uncertainties for the decay constant
$f_K$ \cite{fkvalue}, the theoretical uncertainties for $m_s$ \cite{msvalue}, $m_b$ \cite{mbvalue,narisson}, $f_B$ and $f_{B^*_s}$ \cite{fbsvalue} and the variations over the QCDSR parameters that are the Borel mass, $ M^2 $, and the thresholds $\Delta s $ and $\Delta u $.

\begin{figure}[b] 
\centerline{\epsfig{figure= 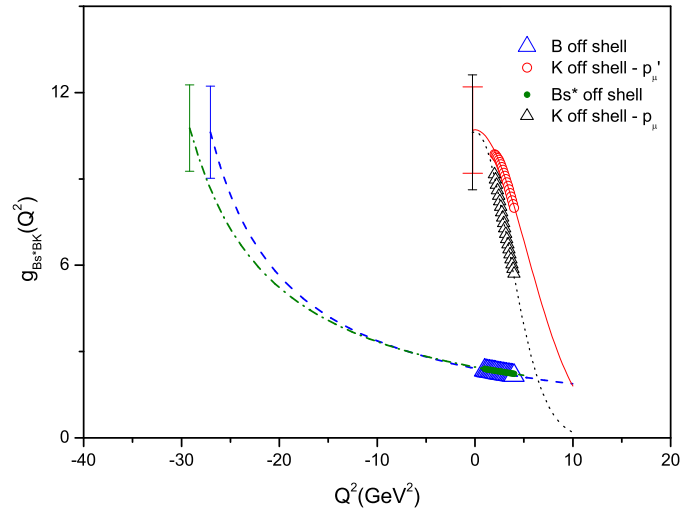,height=80mm}}
\caption{The three cases considered here: off-shell $K$, $B$, $B^*_s$
and for all possible sum rules.}
\label{erros}
%\end{center}
\end{figure}

To compute the error, we proceed in the following way: after calculating
the coupling constant, we compute a new coupling constant, but
now with all parameters kept fixed, except one, which is changed according with its intrinsic error, given in Table \ref{desvios}. Then, we move to the next parameter to be varied, keeping all others fixed. At the end of the procedure, we know how sensitive is this vertex 
respect to each parameter. In table \ref{desvios}, that contain the percentual deviation
for the three cases ($B$, $K$ and $B^*_s$ off-shell mesons), we observe that the biggest variation on the coupling constant is produced when $f_B$, $f_{B^*_s}$ and $m_b$ were varied. 
%---------------------------------------- Tab.(2)-----------------
\begin{table}[h!]
\begin{center}
\begin{tabular}{|c|c|c|c|c|c|}\hline
               &    \multicolumn{3}{|c||} {\textbf{Deviation {\%}}} \\\hline      
\textbf{Parameters} &\textbf{ $B$ off-shell}& \textbf{$K$ off-shell}& \textbf{$B^*_s$ off-shell }\\ \hline
$f_K =159.8 \pm 1.4 \pm 0.44 $ (MeV)    &      $1.04$    & 1.04 &  1.02    \\\hline
  $f_B=208 \pm 10 \pm 29 $ (MeV)  & $15.92$ & $15.97 $ & $ 15.94 $          \\\hline
  $f_{B^*_s}=250 \pm 10 \pm 35 $ (MeV) &  $15.22$ & $15.25$ & $15.29$      \\\hline
  $M^2 \pm 10\% $ (GeV) & $6.05$ & $ 5.9$  & $2.41$                \\\hline
  $m_b=4.20+0.17-0.07$ (GeV)& $13.69$ & $28.80$& $14.18 $              \\\hline
  $m_s=104 + 26 - 34$ (MeV)& $8.98$   &$14.67$ & $5.59$         \\\hline
  $\Delta s \pm 0.1$ e $\Delta u \pm 0.1 $(GeV)  & $13.51$ &$ 2.79$& $14.30$  \\\hline
\end{tabular}
\caption{Percentage deviation related with each parameter. }
\label{desvios}
\end{center}
\end{table}
Finally, the uncertainties for each coupling constant, which is
obtained with $B$, $K$ and $B_s^*$ off-shell are:  
\beq
 g^{(B)}_{B_s^* BK}=10.6 \pm 1.5 
\eeq
for $B$ off-shell meson, 
% $$g^{(K)}_{B_s^* BK}=10.5\pm 1.5$$ for the $ p'$ structure and
%$$g^{(K)}_{B_s^* BK}=10.6 \pm 2.0$$for the $p$ structure. 
\beq
 g^{(K)}_{B_s^* BK}=10.3 \pm 1.7
\eeq
for $K$ off-shell, where were used both structures
that give good SR, and 
\beq
g^{(B_s^*)}_{B_s^* BK}=11.0\pm 1.5
\eeq
for $B^*_s$ off-shell.
Fig.(\ref{erros}) summarize these results, where the error bar 
represents the uncertainties for each calculation.

Considering the three off-shell cases, for all possible and "good" sum rules structures, we obtain the coupling constant of the $B_s^* B K$ vertex. After computing all the uncertainties, we obtain the mean value:
\begin{equation}
g_{B_s^* BK}=10.6 \pm 1.7 
\label{finalcoupling}
\end{equation}
The uncertainties of the QCDSR coming from different sources (other structures with the same mesons of shell, other heavy meson off-shell, the extrapolation fit of the sum rule result and the errors in 
masses, decay constants, condensates, the choice of the Borel mass and the continuum threshold parameters) is near 16 \%. The usual errors in the QCDSR are around 20\% 
\cite{Leinweber}.

Also, from  Eqs.~(\ref{monoBoff}), (\ref{gaussKoff}) and (\ref{monoBsoff}), we can extract the cut-off parameters, $\Lambda$, associated with the form factors. In table \ref{cut-off}, the cut-off parameters are showed and
we can observe that they follow the same trend as we had been observed in Refs.\cite{bclnn01,smnn04}: the value of the cut-off is directly associated with the mass of the off-shell 
meson probing the vertex. A harder form factor means bigger cut-off parameter. 
\begin{table}[h!]
\begin{center}
\begin{tabular}{|c|c|}\hline

  Meson off-shell          & $\Lambda$ parameter $(GeV^2)$ \\\hline
  $ B $               & $34.99$                \\\hline
  $ B^*_s $              & $37.80$               \\\hline
  $ K  $           & $7.49$              \\\hline
\end{tabular}
\caption{Values of the cut-off parameter $\Lambda$.}
\label{cut-off}
\end{center}
\end{table}

Recently, our extrapolation procedure \cite{bclnn01}-\cite{review} was used to calculate the same coupling constant, $ g_{ B^*_{s} B K}$, by the authors of Ref.~\cite{sundu}. Even if the procedure to
obtain the coupling constant was the same, the form factors obtained were different. This calculation do not include the analysis of the pole-continuum contributions, exponential extrapolations functions  with three parameters were used to fit the form factors and different values for the parameters were used in the SR, resulting in a different value of the coupling constant.  

 There are some possibilities that could be considerer to compare our
result with other theoretical predictions.
If we use arguments of heavy hadron chiral perturbation theory (HHChPT), the couplings for the bottom-light vertex $g_{ B^*_s B K}$ are related to the charm-light vertex 
$g_{D^*_s D K} $ through the relation \cite{wise}-\cite{casalbuoni}:  
$$ g_{B^*_{s} B K} = g_{ D^*_s D K} \frac{m_B}{m_D},$$ 
where if it is used $ m_B = 5.279$,  $ m_D=1.8693$ and $g_{D^*_s D K} = 2.84$( which is our QCDSR result \cite{angelo06}), we obtain $ g_{B^*_{s} B K} = 8.02 $, that is in complete
agreement with the result of this paper. 

 Also, $g_{B^*_s B K}$ is related to $g_{B^* B \pi}$ by $SU(3)$ symmetry and
the coupling constants became equal $g_{B^*_s B K}= g_{B^* B \pi}$. 
In table \ref{theoreticalestimatives}, we present other theoretical 
estimatives from QCDSR, Light Cone SR and other methods of the $g_{B^* B \pi}$. 
We find out that $SU(3)$ symmetry present different degrees of
violation where is compare with our estimative of the coupling constant. The third column in the table \ref{theoreticalestimatives} show these results and we obtained similar conclusion in
our review  \cite{review}.
\begin{table}[h!]
\begin{center}
\begin{tabular}{ c c c } \hline
   Approach  & $g_{B^* B K}$ & violation\\\hline
   QCDSR\cite{col} & $20 \pm 4 $ & $ 50\% $                           \\
   QCDSR\cite{col} & $15$       & $ 44\%$                     \\
   LCSR \cite{bel} & $28\pm6$  &$ 65\%$\\
   QCDSR \cite{dn} &  $14\pm4$  &$25\%$\\ 
   LCSR \cite{kho} & $22\pm9$ & $ 55\% $\\ 
   QCDSR \cite{02r} & $42.5\pm2.6$ & $76\% $  \\ 
   QCDSR plus meson loops \cite{mane} & $44.7\pm1.0$ & $73\% $ \\ 
   dispersive quark model\cite{Melik}  & $32\pm5$ &$69\%$ \\
   Dyson-Schwinger equations \cite{bruno}  & $30.0^{+3.2}_{-1.4}$ & $ 73 \%$ \\ 
\hline  
\end{tabular}
\caption{Summary of theoretical estimates for $g_{B^*B\pi}$.}
\label{theoreticalestimatives}
\end{center}
\end{table}

Concluding, we calculate the form factors of the vertex showing that they have two different forms, one when the heavy meson is the off-shell  particle and another when the light meson is the off-shell one, with their extrapolations to the on-shell point leading to the coupling constant. 

In this paper, we made all the calculations taking control of the SR stability and the pole contribution is always contributing more than the continuum. Besides, we computed all the ``good " sum rules and 
we obtained the same value of the coupling constant. Furthermore, we computed the sum rules taking into account the errors in masses, decay constants, condensates, choice of the Borel mass, continuum threshold parameters, obtaining a final result with errors that are near  16\%.

\acknowledgments
This work has been partly supported by the Brazilian funding agencies CAPES and CNPq. We are
deeply grateful to M. Nielsen discussion.

\end{document}